# A USE CASE DRIVEN APPROACH FOR SYSTEM LEVEL TESTING

Muhammad Touseef

muh.touseef@islamabadclub.org.pk

UIIT, PMAS, Arid Agriculture University,

Rawalpindi, Pakistan

Zahid Hussain Qaisar

zahidqaisar@yahoo.com

Assistant Professor, Computer Science Department,

Institute of Engineering and Technology,

National Fertilizers Corporation (NFC-IET)

Multan, Pakistan

## ABSTRACT
Use case scenarios are created during the analysis phase to specify software system's requirements and can also be used for creating system level test cases. Using use cases to get system tests has several benefits including test design at early stages of software development life cycle that reduces over all development cost of the system. Current approaches for system testing using use cases involve functional details and does not include guards as passing criteria i.e. use of class diagram that seem to be difficult at very initial level which lead the need of specification based testing without involving functional details. In this paper, we proposed a technique for system testing directly derived from the specification without involving functional details. We utilize pre and post conditions applied as guards at each level of the use cases that enables us generation of formalized test cases and makes it possible to generate test cases for each flow of the system. We used use case scenarios to generate system level test cases, whereas system sequence diagram is being used to bridge the gap between the test objective and test cases, derived from the specification of the system. Since, a state chart derived from the combination of sequence diagrams can model the entire behavior of the system. Generated test cases can be employed and executed to state chart in order to capture behavior of the system with the state change. All these steps enable us to systematically refine the specification to achieve the goals of system testing at early development stages.

## Categories and Subject Descriptors
Requirement validation, System Testing, State based testing, Technique and method for system testing by requirement analysis.

## 1. INTRODUCTION
Systematic testing is the only key process to accomplish higher quality software. The step wise refinement model for software testing is proposed to achieve high quality software that can be achieved by refinement of system requirements which serves as a strong basis for system testing.

Generally user requirements are stated in terms of use case scenarios that describe user needs relating with the system behavior in the form of user-system interaction showing system behavior in operation. Initially, Informal set of user requirements are used to satisfy and derive scenarios; A Use case scenario describes detailed description of one specific usage or the specification of that part of the system. Analysis of use case scenarios provides a complete understanding of the system [2]. Which are then transformed to semi formal model using graphical notations such as use case, this semi formal model is source to derive system level test cases, as it defines major system components and interactions among them. Use case based testing deals with generation of test cases from the system requirements. These test cases are then exercised to show that the system conform its specification and its overall behavior is accurate. Hence, use cases provide a foundation for the system level testing [10]. The basic principle behind the system testing is to verify the functional and performance aspects of the intended system [2].

A lot of research work is reported in the literature on use case and scenario based system testing. The most important work on the topic is of Briand et.al [2], Nebut [9] and Whittle [15]. They present the system testing using use cases. The major limitation of their proposed work is absence of formalized test case generation based on control flow with guards. Hence formalized test case generation based on control flow by passing each of the guard is not available yet. Similarly Sequence Diagram and a State Chart can be used for system behavior validation. In the proposed approach guards are added to the use cases that help to capture the sequential events alternatively. In our approach test requirements are generated as logical expressions with the help of contracts discovering the path flow. A refinement however, is required to know behavior of each system component in more precise, concrete and formal manner.

The rest of the organization of the paper is as under Section 2 consists of related work of use case based modeling and testing techniques. In section 3, we have discussed our proposed approach with results and discussions section. Section 4 describes the proposed solution with the help of some example section 5 presents the conclusion.

## 2. Related Work
In this section we are going to discuss the related work in the domain of use case based testing.

Regnell [12] provided a method of creating a synthesized use case model. Ryser and Glinz [13] presents a technique for the description of use cases with scenarios showing the flow of events with pre- and post-conditions (system states) for the use case, which is a formal representation of the flow of events.

The most important aspect of use case based testing is the generation of test cases at the early stages which helps in refining unclear and poorly defined requirements Blackburn [18]. By eliminating model defects before the coding begins and the test case creation results in significant cost savings and higher quality code because the later the defects captured they are more costly both in effort and time.

A use case based testing deal with capturing of user requirements and the generation of test cases for the system at early stage in the engineering process and validating the tests with the specification of the system. Many approaches have been cited in the literature. Major work can be found in Briand and Labiche [2] that involve use case diagram, activity diagram and sequence diagram that for the generation of system tests cases. Use case dependencies are modeled by an activity diagram and the class diagram is used to show the functionality of the system. Testability requirements are generated from the sequential constraints between the use cases described in meta-model which include formal description of class, attributes, operators and contracts. Nebut [9] enhances the Briand and Labiche [2] work with the introduction of contracts. Kim [6] discusses application of the state diagram in UML to class testing where test cases are generated by using either flow control or data flow technique. Raza [10] proposes a test path generation approach for scenarios by applying coverage criteria.

Hsia [4] Describe user oriented scenario trees that represent all scenarios for a particular user. A scenario tree consists of state nodes and event directed arcs, Regular expressions are used to formally state the user scenario that results in a deterministic finite state machine with a single state that defines both its initial and terminal state. Kosters [7] present an approach for mapping use cases onto static classes and methods. The technique transforms the scenario steps into actions. Use case expansion is described by directed use case graph where nodes inherits the scenario each scenario step is developed by method of tree. Whittle [17] mainly focus on the generation of hierarchical state machines through a synthesize algorithm that transforms scenarios into state machines deriving from use case charts. Alspaugh [1] presents goal/requirement based V&V in order to develop requirement scenario description language "ScenarioML" used to generate functional requirement goals. The "goals and Intentions" verification helps in distinction of false claims while goal establishment provide more confidence of testing with less effort and hence cost-effectiveness is improved. The scenarios and use cases go until goal success or abandonment. Test case generation can be done using test coverage metric that can be to cover all the sub-goals in the event tree and the test suit consist of set of event traces that integrally provide requirement goal coverage. Briand and Labiche [2] Proposed an approach for system testing by comparing system behavioral aspects with specifications and ensuring the system behaves as required and describe in the specification. They had used UML analysis artifacts to derive system test requirements which require execution of test scenarios with specification. Nebut [9] proposes approach inspiration of Briand and Labiche [2] work UML based approach to system testing. Contract language for requirements is defined as pre and post conditions associated as logical expression. Regnell and Runeson [11] proposed a synthesis phase extension to the OOSE use case modeling approach. In their approach, separate use cases are integrated into a synthesized usage model. The synthesis phase consists of three activities; formalization, integration, and verification. Usage testing through automatic generation of test cases is derived from the usage views. Kim [6] discusses application of state diagram in UML to class testing by proposing a set of coverage criteria based on control and data flow in the UML state diagram. The set of states represents both the basic and composite states which contain other states as sub states and are defined as either OR-State or AND-State. States can have actions associated with them that contains list of operations for transition being occur. Test cases are generated by either using control flow or data flow technique. Raza [10] proposes a test path generation approach for scenarios using the interaction overview diagram "IOD" to express the scenarios. Yang Liu and Yafen Li [20] proposed a technique for test case generation using model based architecture. J.J Gutierrez et al [5] proposed technique for test generation using model based architecture. Patrícia Machado [11] has proposed similar kind of approach. A contract transition system "CTS" is build from the operational contracts in the IOD that specify the pre and post conditions. The approach identifies operations in the IOD and then the CTS matrix is developed that identifies states and contracts for the CTS. For the generation of contract transition system CTS scenarios are identified from the IOD for individual use case and are represented as CTS based on the CTS matrix. Test paths can be create by applying coverage criteria i.e. all transition coverage or all state coverage.

Most of the approaches present in the literature involve more functional details i.e. Briand and Labiche [2] uses a class diagram which require more functional analysis of the system that can be difficult very early in the design phase. Whittle and Praveen [17] mainly focus on the generation of hierarchical state machines by describing a synthesize algorithm that transforms scenarios into state machines without applying guard. Use case scenario is created for each use case, from each use case scenario node sequence diagram is generated and finally by the combination of sequence diagram a hierarchical state chart is generated without applying any guards and hence testing criteria and testing is not consideration. Briand and Labiche [2] derive system test requirements using of UML analysis artifacts, system test requirements are generated from the Meta model based on the sequence diagram that describes each class, method and attribute. Nebut [9] inspiration of Briand's work presents UML based approach to system testing by defining Contract language for requirements as pre and post conditions associated as logical expression. We have presented an approach that has inspiration from Briand and Labiche [2], Nebut [9] and Whittle [17] work our approaches differs with the fact that we are taking into account only the specification of the system without involving the functional details so a level above on the specification by capturing the sequential ordering of the use cases with the guard annotation defined as contracts. Addition of Contracts in the proposed approach is closer to the way Nebut [9] applied the contracts to use cases whereas Briand and Labiche [2] and Whittle [17] does not imposed contracts. The proposed approach applied contracts on the use cases to capture the sequential dependencies and the annotation of contracts on the use case scenario is used to generate the test objectives whereas Nebut [9] does not imposed contracts on use case scenarios furthermore test objectives are created based on the coverage criteria. The advantage of generating test objectives from contracts makes them executable by defining as logical expression. The proposed approach captures use case flow model and contracts from the specification. With the addition of guards at the use case sequential flow allows tracking the path selection at the top node, with the introduction of guards to each use case enables to strength the conditional execution flow of use cases. Introduction of guards to the scenarios enables to make a conditional testing likewise generation of conditional test path selection becomes easy and can be defined as logical expression. The sequence diagram generation by the proposed approach from use case scenario which is more appropriate if guards are available at the use case scenarios, where the guards of the use case scenario becomes messages for the sequence diagram. The sequence diagram can be generated from the use case scenario by using Whittle [17] synthesis algorithm but the proposed approach generates a Separate contractual sequence diagram is generated for each alternative sequence of a use case scenario with the contract extraction from the use case scenario to sequence diagram. Similarly from use case sequence to state chart provide a complete conditional flow that makes easy to test the system behavior against events, hence enhances the power of testing at the analysis level. In the proposed approach we are generating contractual state chart by the combination of sequence diagrams through an algorithm which creates a state chart transition table. Whittle [17] also proposed a synthesis algorithm for state chart generation but does not imposed contracts whereas Briand and Labiche [2] and Whittle [17] uses Activity diagram instead of state chart.

Our contribution to literature is the extraction of sequential dependencies of use cases involving use cases contracts and extraction of test objectives from the use case scenario contracts both expressed as logical expression furthermore generation of contractual state chart.

## 3. Proposed Approach

In this section, we discuss our proposed approach for system testing. Our approach is inspired from Briand and Labiche [2], Nebut [9] and Whittle [15] work. Our approache differs with the fact that we are taking into account only the specification of the system. The proposed approach does not involve functional details so a level above on the specification by capturing the sequential ordering of the use cases with the guard annotation defined as contracts. Addition of Contracts in the proposed approach is closer to the way Nebut [9] applied the contracts to use cases whereas Briand and Labiche [2] and Whittle [15] does not imposed contracts. Our proposed approach applies contracts on the use cases to capture the sequential dependencies and the annotation of contracts on the use case scenario is used to generate the test objectives whereas Nebut [9] does not imposed contracts on use case scenarios furthermore test objectives are created based on the coverage criteria.

In this section we are going to discuss our proposed approach for system level testing based on scenarios. Our technique uses following steps

1. Overall System use case diagram
2. Generation of Sequential Use Case Diagram
3. Extracting Sequential Constructs for use cases from specification
4. Deriving the second level use case scenario diagrams where each node express/explores the level-1 use case node with guards
5. Generating execution contracts to level 2 scenario use cases as logical expression
6. Extraction of test Goals from Contracts
7. Deriving the contractual Sequence diagrams from use case scenario
8. State Chart Transition Table Generation from combination of sequence diagrams
9. Generation of contractual state diagram from state chart transition table
10. Test Goals Execution on state chart

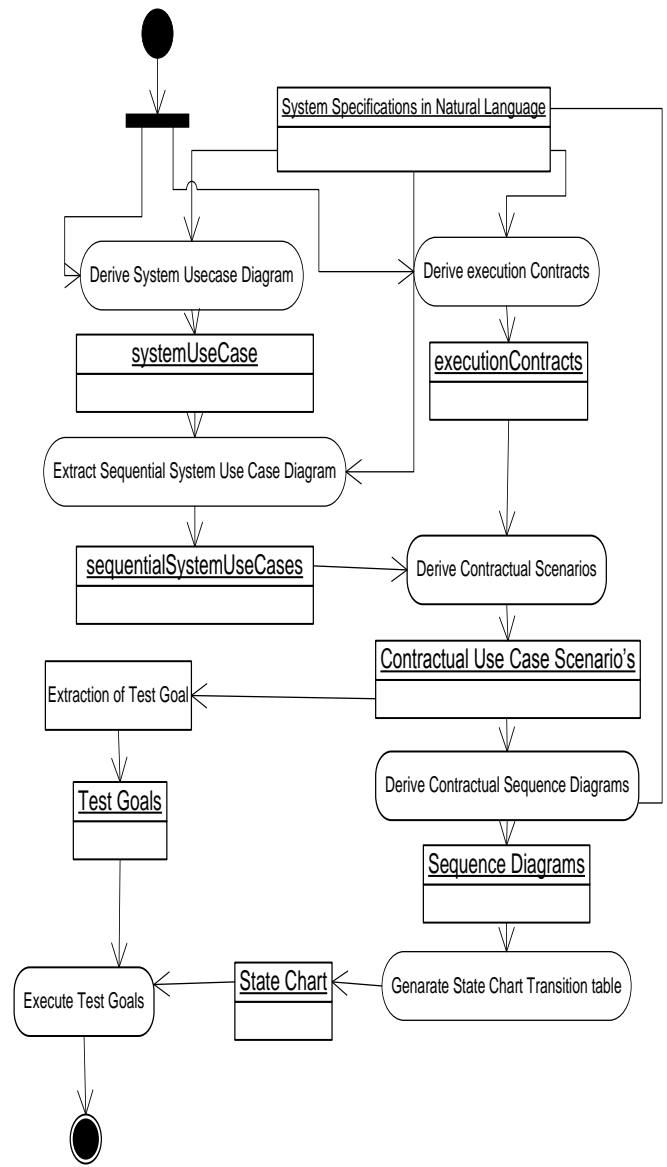

**Figure 1: Abstract Flow Model of Proposed Approach**

### 3.1 Overall system use case design diagram
The use case design diagram represents the entire system usage where nodes are use cases. The number of use cases may be very large in the system. Each of the use case contains its own set of events to occur, therefore the entire system use case diagram can comprises of several use case nodes by involving interacting actors [9].

### 3.2 Generation of sequential use case diagram
A use case based requirement validation requires that the sequential ordering of the use cases should be captured in behavioral model and can be the first component of system test requirements [2]. The use case sequential flow describes how the use cases track each other and gives a clear idea of system usage [16].

### 3.3 Extracting sequential constraints for use cases
The sequential constraints between the use cases can be specified by using the logical expression with the AND/OR operators, where the OR operator show the alternative paths in the execution order [16]. As we are adding contracts to the use cases so the sequential contracts will be made with the combination of guards/contracts.

### 3.4 Generation of use case scenario diagram
A use case scenario is a system usage view of a specific actor which can be a user, external system or communicating device [9]. Use cases scenario normally focus on the behavior of the system and typically describe several paths for a use case and simulate the sequence of actions to real happenings as expected to occur when the system is in operation [8]. We are generating scenario chart from the specification of the system with addition of guards to the scenario nodes.. Addition of guards allows the requirement validation and test case generation [9].

### 3.5 Generating execution contracts
The execution contracts are generated from the use case scenario guards applied to the sequential constructs, where as the alternative path are covered by oring the decision conditions.

### 3.6 Test goal extraction
Test goal specifies the objective for test i.e. what the user or tester require from the system should be identified separately. Identification of goals gives confidence to testing, the goal plan should include the alternatives as well [1].The primary advantage of using contracts is the definition of test goals but these should be consistent while moving from one stage to other in order to make consistent and proper execution of test goals [9]. Test goals are extracted from the execution contracts for each of the alternative a separate test goal has been identified.

### 3.7 Deriving the contractual system sequence diagrams
Sequence diagram shows the sequence of events as appeared in the scenario with one nominal and number of exceptional scenario involving the system and the participating actor.

Sequence diagram contains more information than the use case scenario while at the same time use case scenario contains more information about pre and post conditions [9]. Hence sequence diagram can be used to bridge the gap between the test objectives and test cases alternatively depicting the use case scenario [9].

### 3.8 State chart transition table generation from combination of sequence diagrams

The state chart transition table is created from the combination of sequence diagrams, as each sequence diagram consider the message state from where the system gets the message. So it helps to easily translate the sequence diagrams into state chart with the help of transition table. For the generation of transition table we are introducing an algorithm.

**Table 1**: **Algorithm to Generate a State Chart Transition Table**

*Input*. Combination of sequence diagrams belongs to a single use case scenario

*Output. A State Chart Transition Table* with 5 columns:

Column 1:        Contains State

Column 2: Contains Guard to move the alternate State

Column 3: Contains Next State By Passing the Guard

Coulmn 4: Contains alternative State

Column 5: Contains Guard to Reach the alternative State

1 Algorithm Generate_State_Chart_Transition_Table
2 Body
3    GenerateSeqNodes
4    SortSeqNodes
5    Generate StateTable
6 End
1 Function GenerateStateTable
2     Var i := 1 ∈ Number
3     Var StatTab[i][5] ∈ TwoDArray
4     SortSeqNode := 1st(SortSeqNode)
5     While (SortSeqNode not end) do
6         If SortSeqNode[Previous] = null then
7             SortSeqNode = Next SortSeqNode
8         End If
9          StatTab[i][1]:=SortSeqNode[Previous]
10         StatTab[i][2]:=SortSeqNode[Guard]
11         StatTab[i][3]:=SortSeqNode[State]
12     SortSeqNode = Next SortSeqNode
13         If  SortSeqNode[Previous] = StatTab[i][1]
14         StatTab[i][4]:=SortSeqNode[State]
15         StatTab[i][5]:=SortSeqNode[Guard]
16       End If
17      SortSeqNode = Next SortSeqNode
18      i++
19    done
20 End

### 3.9 Generation of contractual state diagram from state chart transition table

Since each of the alternative sequence is described independently with its own specific order of events, by these can cause in the introduction of inconsistencies that must be detected and resolved. UML sequence diagram can not contain enough details for the detection and resolution of such conflicts. State charts; models the system behavior against the events and can be helpful for resolving them [15]. We are generating state chart from the transition table that is inherited from the combination of sequence diagram so sequence diagram messages will be converted into guards to the state chart making the execution of state chart with contracts, and making possibility for the execution of test goals defined earlier at the use case scenario description in the form of test goals.

## 4. Case Study

For the case study we are using Inventory System.

### 4.1 System Specification

1. Only authorized user can access the system
2. The first step will be to create a Purchase Requisition for the item indicating the item required
3. Purchase order for an item can be made only for the completed Purchase Requisition
4. Purchase order can be put to registered vendor against the requisition
5. The item for which there is purchase order must be stocked in the system
6. A Store Requisition for the issuance of item can be made possible only if the item is stocked in
7. A stock out can be made for an item only against the store requisition

#### 4.1.1 Overall system use case design diagram

Figure 2 represents the entire system use case where the actors that are interacted to the system are defined.

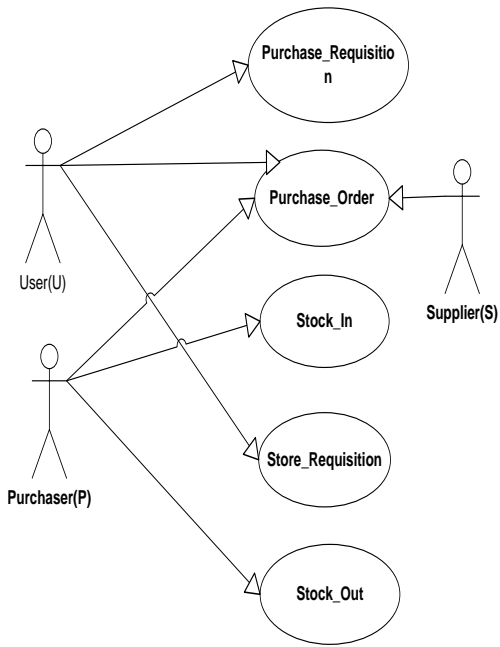

**Figure 2 Entire System Use cas**

#### 4.1.2 Generation of sequential use case diagram

Fig 3 shows an entire sequential use case with guards applied; the entire sequential use case shows the execution flow of the whole life cycle of the system with Pre and Post Condition of each use case representing a use case node.

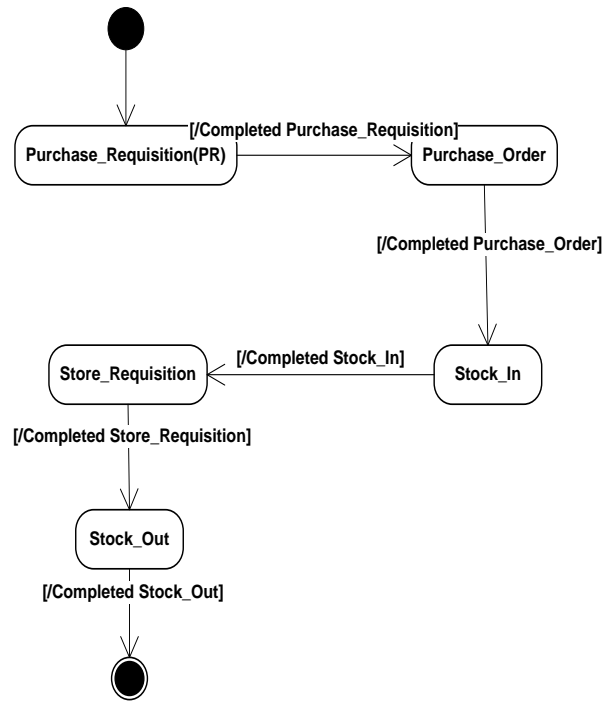

**Figure 3 Sequential System Use case diagram**

#### 4.1.3 Extracting sequential constraints for use cases

The sequential contracts for the entire system use case is derived by following the path in the transition as logical expression by using the "AND/OR" logical operators. Where OR indicates optional path of the system flow.

**[/Completed Purchase_Requisition and /Completed Purchase_Order and /Completed Stock_In and /Completed Store_Requisition and /Completed Stock_Out]**

For extraction of Sequential Contracts each of the use case nodes i.e. used in fig 2 has to be involving path execution of the whole system.

#### 4.1.4 Generation of use case scenario diagram

For each of the use case there will be a scenario indicating the ordering of events in the use described as use case scenario. As there are multiple use cases in the system so for each use case there will be a separate scenario diagram. We are only dealing with the use case scenario Purchase Requisition (PR) here.

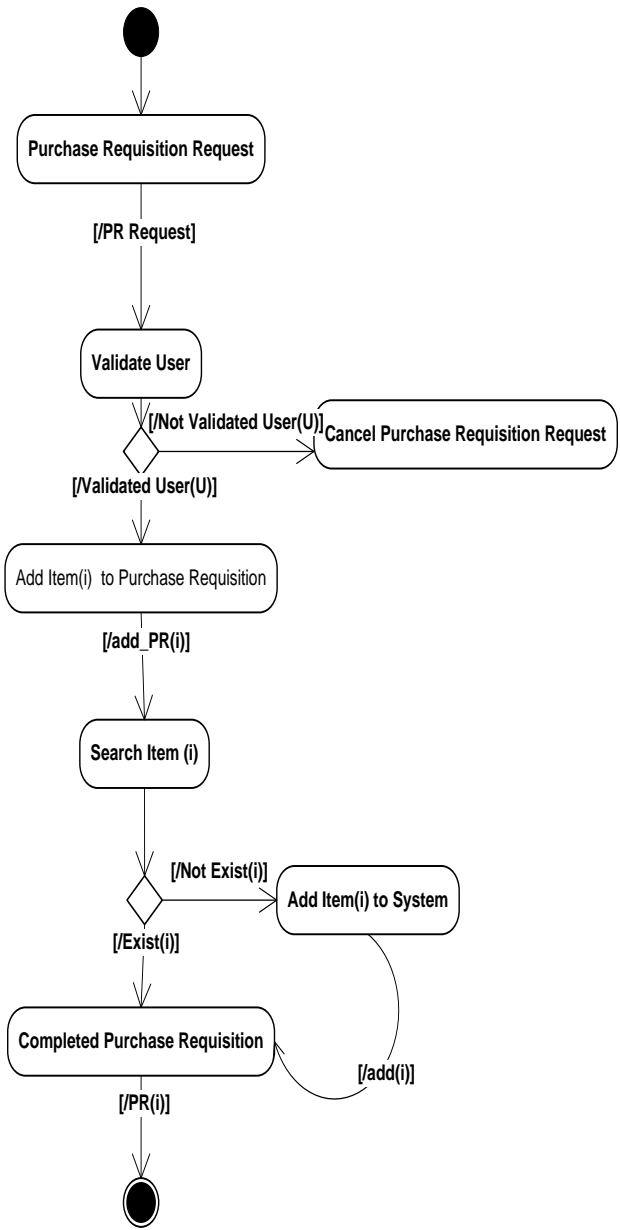

**Fig 4 Use Case Scenario for Purchase Requisition**

#### 4.1.5   Generating execution contracts

Contracts generated from the use case scenario will be used to define the test goals by routing through the path.

**Pre Condition:**   User(u)

**Execution Contracts:**   [/PR_Request and {(/Validated User (U) and /add_PR(i) and (exist(i) or (Not /Exist(i) and /add(i))) and /PR(i)) or /Not Validated User(U)}]

**Post Condition:**   PR(i)

#### 4.1.6   Test goal extraction

Test goals are extracted from the execution contracts defining the path flow for the scenario. Each test goal defines the alternative path of the scenario.

**Test Goal TG_PR1**

   TG_PR1= [/PR_Request and /Validated User (U) and /add_PR(i) and exist(i) and /PR(i)]

**Test Goal TG_PR2**

   TG_PR2= [/PR_Request and /Validated User (U) and /add_PR(i)  and Not /Exist(i) and /add(i) ) and /PR(i)]

**Test Goal TG_PR3**

   TG_PR1= [/PR_Request and //Not Validated User (U)]

#### 4.1.7   Deriving contractual system sequence diagrams

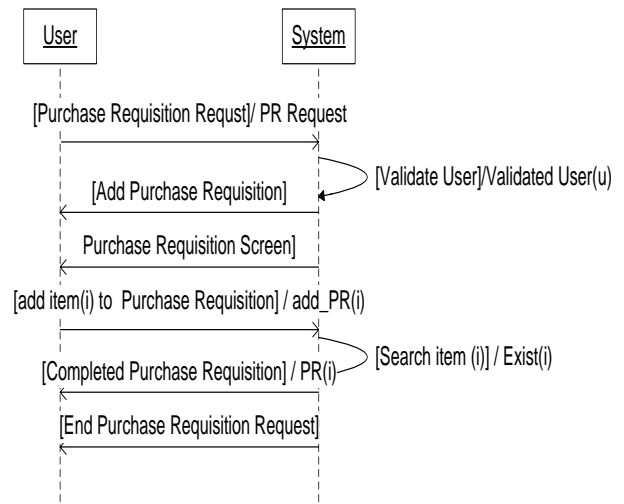

**Fig 5      Sequence Diagram for Purchase Requisition**

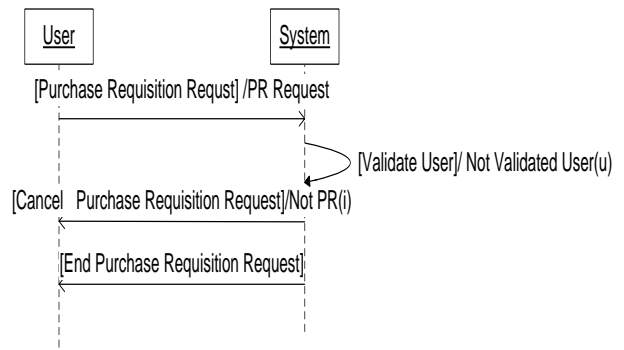

**Fig 6      Sequence Diagram 2 for Purchase Requisition**

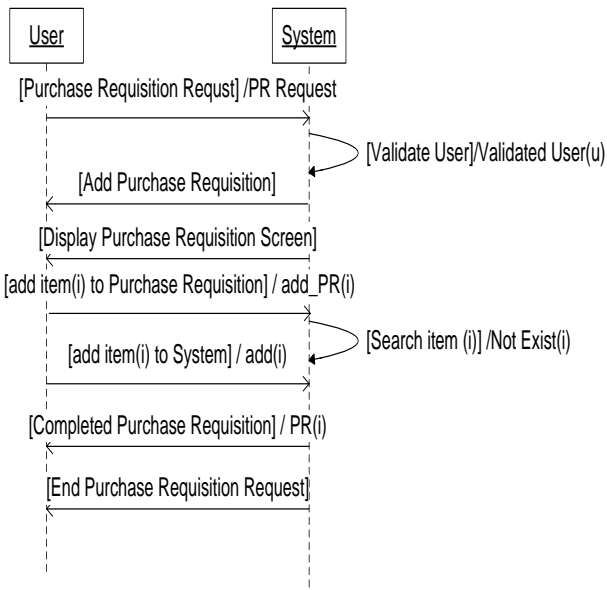

### 4.1.8 State chart transition table generation from combination of sequence diagrams

State Chart Transition table is generated from the Algorithm defined. The Transition Table contains five Columns State, Guard, New state after passing the guard, alternative state defines if the guard condition does not satisfy then the alternative route should be adopt where the alternative state guard is the passing condition for the alternative state respectively. The state chart transition table generated from sequence diagram 1, 2 and 3 are as follows.

**Fig 7     Sequence Diagram 3 for Purchase Requisition**

**Table 2: State chart transition tables for PR**

| State | Guard | New State | Alternative State | Alternative State Guard |
|---|---|---|---|---|
| Purchase Requisition Request | /PR Request | Validate User | | |
| Validate User | /Validated User(U) | Add Purchase Requisition | Cancel PR Request | /Not Validated User(U) |
| Add Purchase Requisition | | Display Purchase Requisition Screen | | |
| Display Purchase Requisition Screen | | Add item (i) to Purchase Requisition | | |
| Add item (i) to Purchase Requisition | /Add_PR(i) | Search item (i) | | |
| Search item (i) | /Exist (i) | Completed Purchase Requisition | Add item (i) to System | /Not Exist (i) |
| Completed Purchase Requisition | /PR(i) | End Purchase Requisition Request | | |
| Cancel Purchase Requisition Request | /Not PR(i) | End Purchase Requisition Request | | |
| Add item (i) to System | /Add(i) | Completed Purchase Requisition | | |

### 4.1.9 Generation of contractual state diagram from state chart transition table

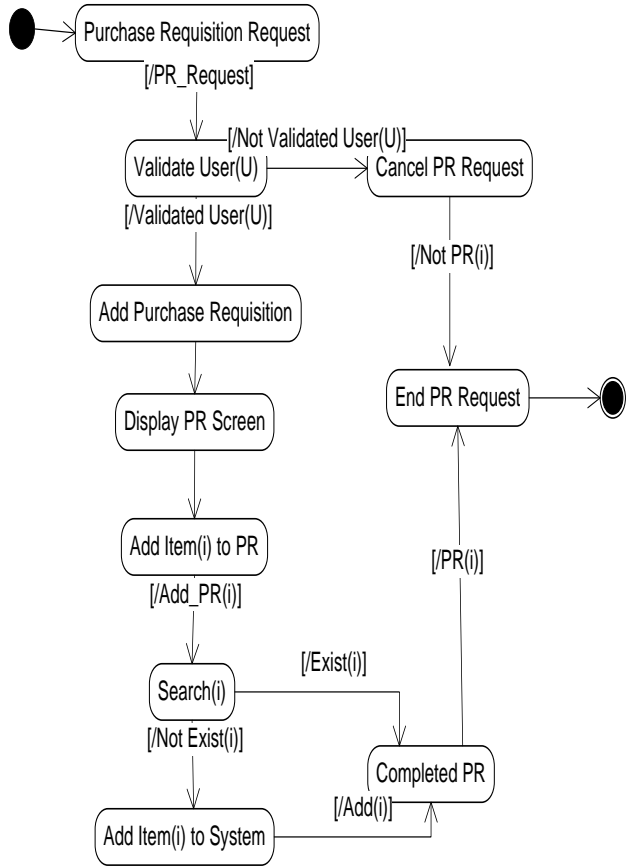

**Fig 8        State Diagram 3 for Purchase Requisition**

## 5. Results and Discussion

We are generating results based on the related techniques that presents use case based system testing. [2] work provide a base for system testing based on use cases [9] extends by adding contracts. However lack of some formalization technique for properly test case generation and to maintain consistency between use cases to scenario and state chart generation is sensitive issue. The main advantages of the proposed approach as under

### 5.1 Use Case Sequential Ordering
Addition of guards to the use cases as pre and post condition enables to formally express sequential flow as logical expression. AND/OR logical operators can be used to identify execution paths where the OR logical operator shows the alternative paths in the system. The advantage of current proposed approach is that it allows the addition of guards to use cases which added more strength to testing by aiding to generate complete test conditions with guards and enabling to derive conditional test case generation also sequential flow can be tested by guards easily.

### 5.2 Contractual Use Case Scenarios
A Use Case Scenario presents the execution trace of a system and provides a base for the development of state machine [2]. Use case scenarios can be expressed by using the sequence diagram that shows the flow of events [2] but it is difficult to define guards at the sequence diagram. However through pre and post conditions applied to use case scenario enables the generation of test paths. The proposed approach also applies the contractual sequence diagram derived from the use case scenario that can be used to bridge the gap between the test objectives and test cases alternatively depicting the use case scenario as it may contain additional information than scenario.

### 5.3 Test Goal Generation through Scenarios
The advantage of applying guards at the scenario enables to generate the test cases also referred to as test goals. These test goals capture the flow of events for the use case scenario. As the test goals are based on contracts so that can be formalized as logical expression.

### 5.4 Contractual State Chart
State diagrams represent the object behavior with invocation of event "represent operation" and are used to record different states with events that can cause a state transition. A state machine is composed of state representing the behavior of the system on certain input whereas transition may result in an output action, event "an input" and action the output result [17]. State diagram annotation with guards "Guards are associated with pre and post conditions" enables to specify the entry and exit conditions. Optional Guards can be added to states and transition may be annotated with guard, event, and action. If there is no guard or both guards are true then the exit action is performed. Test cases are imposed to verify the behavior of the system when applied on the state chart.

We had implemented a tool that takes XML containing guards of scenario as input and generate test path expressing test cases as logical expression.

## 6. Conclusions
In this paper, we presented a scenario based testing technique for system level testing. The main aim of the proposed approach is to generate formalize test cases by applying guards on the scenarios covering conditional flow path coverage criteria. For every use case scenario in the system, we generated a sequence diagram by utilizing the guard conditions on the scenarios. By combining the generated sequence diagrams, we generated a state chart depicting the overall behavior of the system.

We applied the proposed approach on an inventory system case study. We created an entire system level use cases then sequential use case diagrams is generated through the contracts showing the whole system execution path. The advantage of applying contracts at the scenario enables to generate the test cases / test goals and enables us to validate system from the user as at this point user can view what are the actual steps involved in the system usage.